\documentclass[10pt,a4paper,twocolumn,english,prb,aps,showpacs,preprintnumbers,amsmath,amssymb]{revtex4}
\usepackage{graphicx}
\usepackage[latin9]{inputenc}
\usepackage{url}
\usepackage[update,prepend]{epstopdf}
\usepackage{color}
\usepackage{babel}
\usepackage{amsmath}
\usepackage{amssymb}
\usepackage[unicode=true,
 bookmarks=true,bookmarksnumbered=false,bookmarksopen=false,
 breaklinks=false,pdfborder={0 0 1},backref=false,colorlinks=true]
 {hyperref}
\hypersetup{pdftitle={High temperature superconductivity from realistic Coulomb and Fr\"ohlich interactions},
 pdfauthor={A. S. Alexandrov, J. H. Samson, G. Sica},
 pdfkeywords={High temperature superconductivity, polaron, bipolaron, EPI interaction},
 linkcolor=blue,urlcolor=blue}

\makeatletter

\usepackage{dcolumn}% Align table columns on decimal point
\usepackage{bm}% bold math

\makeatother

\begin{document}

\title{High temperature superconductivity from realistic Coulomb and Fr\"ohlich interactions}

\author{A. S. Alexandrov$^{1,2}$}
\author{J. H. Samson$^{2}$}
\author{G. Sica$^{2,3}$}

\affiliation{$^{1}$Instituto de Fisica ``Gleb Wataghin'', Universidade Estadual de Campinas, UNICAMP 13083-970, Campinas, S\~{a}o Paulo, Brasil\\
$^{2}$ Department of Physics, Loughborough University, Loughborough LE11 3TU, United Kingdom\\
$^{3}$ Dipartimento di Fisica ``E.R. Caianiello'', Universit\`{a} degli Studi di Salerno, I-84084 Fisciano (SA), Italy}

\date{\today}

\begin{abstract}
In the last  years ample experimental evidence has shown that charge carriers in high-temperature superconductors are strongly correlated but also coupled with lattice vibrations (phonons), signaling that the true origin of high-T$_c$ superconductivity can only be found in a proper combination of Coulomb and electron-phonon interactions. On this basis, we propose and study a  model for high-T$_c$ superconductivity, which accounts for realistic Coulomb repulsion, strong electron-phonon (Fr\"{o}hlich) interaction and residual on-site (Hubbard $\tilde{U}$) correlations without any ad-hoc assumptions on their relative strength and interaction range. In the framework of this model, which exhibits a phase transition to a superconducting state with a critical temperature T$_c$ well in excess of $100$K, we emphasize the role of $\tilde{U}$ as the driving parameter for a BEC/BCS crossover. Our model lays a microscopic foundation for the polaron-bipolaron theory of superconductivity. We argue that the high-T$_c$ phenomenon originates in competing  Coulomb and Fr\"{o}hlich interactions beyond the conventional BCS description.
\end{abstract}

%74.20.-z Theories and models of superconducting state
%74.72.-h Cuprate superconductors
%71.38.-k Polarons and electron-phonon interactions

\maketitle

\section{Introduction}

Unconventional symmetries of the order parameter allowed some researchers to maintain that a purely repulsive interaction between electrons (Hubbard $U$) accounts for superconductivity without phonons in a number of high-temperature superconductors \cite{pla}. However, recent analytical \cite{alekab2011} and numerical (Monte-Carlo) \cite{MC,MCHardy} studies shed doubts on the possibility of high temperature superconductivity from repulsive interactions only.

Also a growing number of experimental and theoretical results suggest that strong electron correlations and significant electron-phonon interaction (EPI)  are the unavoidable features for a microscopic theory of high T$_c$ superconductivity \onlinecite{aledev,alespecial}. In particular the doping dependent oxygen-isotope effects  on the critical
temperature $T_c$ and
on the in-plane supercarrier mass
(ref.~\cite{ZhaoYBCO95,ZhaoLSCO95,ZhaoNature97,ZhaoJPCM98,Zhao01,Keller1}),
provide direct evidence for a significant EPI and bipolaronic carriers \cite{asazhaoiso} in high-temperature cuprate superconductors.
Angle-resolved photoemission spectra (ARPES)
\cite{lanzara,shen} provide further evidence for the strong EPI
apparently with c-axis-polarised optical phonons \cite{meev}. Some   theoretical models  show that  detailed understanding of ARPES requires EPI \cite{alechris,mish2,hoh,gun,shen} and the lattice disorder \cite{alekim,ale} to be taken into account along with strong correlations.
These results as well as neutron scattering
\cite{ega,rez}, tunneling
 \cite{Zhao07,Boz08,ZhaoPRL09},  pump-probe \cite{Gadermaier2010}, earlier \cite{mih} and more recent \cite{mish} optical specroscopies  unambiguously show that lattice vibrations  play a
significant but unconventional role in high-temperature
superconductivity.

Since first proposed \cite{AlexandrovRanninger}, much attention has been paid to the strong EPI as a mechanism of superconductivity providing effective on-site and inter-site attractions between small polarons (electrons dressed by a cloud of phonons) \cite{AlexandrovMott}. In the framework of negative Hubbard-$U$ and extended negative Hubbard-U models, the strong electron-phonon coupling results in a bound state of two polarons that condense with a Bose-Einstein critical temperature strictly related to the mobility of the pairs \cite{Loktev}. However, the failure of these models in predicting a high critical temperature, due to localization of pairs in the strong coupling regime with some particular (Holstein) EPI, led to a better understanding of more realistic EPIs as the  competing interactions with respect to Coulomb repulsion.

Analytical and numerical calculations \cite{alebra2,bau} clarified that EPI with high-frequency optical phonons in ionic solids remains poorly screened signaling the presence of long-range (Fr\"{o}hlich) EPI at any density of polarons with a remarkable reduction of the polaron effective mass. Consistently, studies on the so called ``Fr\"{o}hlich-Coulomb'' model (FCM) \cite{FCM,SCinFCM}, in which strong long-range EPI and long-range Coulomb repulsion are treated on equal footing,  predict light polarons and bipolarons (bound state of two polarons) in cuprates with a remarkably high superconducting critical temperature in the range in which all the interactions are strong compared with the kinetic energy of carriers. The interpretation of the optical spectra of
high-T$_c$ materials as the polaron absorption \cite{mih,dev,aledev}
strengthens the view \cite{FCM} that the Fr\"{o}hlich EPI is
important in those compounds.

In most analytical and numerical models of high-temperature superconductivity, proposed so far, one or both Coulomb and electron-phonon interactions have been introduced as input parameters not directly related
to the material. Different from those studies an analytical multi-polaron model of high-temperature superconductivity in highly polarisable ionic lattices has been recently proposed \cite{Alexandrov2011} and numerically studied (for two-particle states) \cite{sica}   with generic (bare)  Coulomb and Fr\"ohlich interactions  avoiding any ad-hoc assumptions on their range and relative magnitude. It has been shown  that the generic Hamiltonian comprising any-range Coulomb repulsion and the Fr\"ohlich EPI can be reduced to a short-range $t-J_{p}$ model at very large lattice dielectric constant, $\epsilon_{0}\rightarrow \infty$, for the moderate and strong EPI. In this limit the bare static Coulomb repulsion and EPI negate each other  giving rise to a novel physics described by the polaronic $t-J_{p}$ model with a short-range polaronic spin-exchange $J_p$ of phononic origin \cite{Alexandrov2011}.

The cancelation of the bare Coulomb repulsion by the Fr\"ohlich EPI is accurate up to $1/\epsilon_0$ corrections.  At finite $\epsilon_0$  a residual on-site repulsion  of polarons, $\tilde{U}$, could be substantial if the size of the Wannier (atomic) orbitals is small enough. Here we study the effect of this on-site repulsion on the ground state of the extended $t$-$J_p$-$\tilde{U}$ model accounting for all essential correlations in high-temperature superconductors. It is worth emphasizing that the effect of the on-site $\tilde{U}$ does not follows as a mere generalization of the $t$-$J_p$ model. The residual Hubbard $\tilde{U}$ in fact leads not only to the suppression of on-site pairs but also to the reduction of the exchange interaction and to the Bose-Einstein condensation (BEC) to BCS  (BEC/BCS) crossover.

\section{Bare Hamiltonian}
Keeping  major terms in both interactions, diagonal  with respect to sites, yield our generic Hamiltonian in the site representation,
\begin{eqnarray}
H &=-&\sum_{i,j} (T_{ij}\delta_{\sigma \sigma^\prime}+\mu \delta _{ij}) c_{i}^{\dagger
}c_{j} +{\frac{\alpha}{2\epsilon_{\infty}}}\bar{\sum_{i\neq j}}{\frac{\hat{n}_{i}\hat{n}_j} {|\mathbf{%
m-n}|}}+\cr && \sum_{\mathbf{q},\nu,i}\hbar \omega _{{\bf q}}\hat{%
n}_{i}\left[ u_i(\mathbf{q} ) d_{{\bf q} }+H.c.\right]+H_{ph}.
\label{genhamiltonian}
\end{eqnarray}
Here
$T_{ij}$
is the bare hopping integral, if $\mathbf{m}\neq \mathbf{n}$, or the site energy, if $\mathbf{m}= \mathbf{n}$, $\mu$ is
the chemical potential, $i=(\mathbf{m},\sigma)$ and $j=(\mathbf{n},\sigma^{\prime})$
include both site $(\mathbf{m,n})$ and spin $(\sigma,\sigma^{\prime })$ quantum numbers, $c_{i}, d_{%
\mathbf{q} }$ are electron and phonon operators respectively, $\hat{n}%
_{i}=c^\dagger_i c_i$ is a site occupation operator,  $\alpha=e^2/4\pi \epsilon_{00}$ ($\epsilon_{00}\approx 8.85 \times 10^{-12}$ F/m is the vacuum permittivity), and $H_{ph}= \sum _{\bf q} \hbar \omega _{{\bf q}} (d_{{\bf q} }^\dagger d_{{\bf q} }+1/2)$ with the phonon frequency $\omega_{\bf q}$.

The EPI matrix element is
\begin{equation}
u_i(\mathbf{q})= (2N)^{-1/2}\gamma(\mathbf{q})\exp(i \mathbf{q \cdot m})
\end{equation}
 with the dimensionless EPI coupling, $\gamma(\mathbf{q})$ ($N$ is the number of unit cells ).
 Deriving the generic Hamiltonian  in the site representation \cite{Alexandrov2011} we  approximate the Wannier orbitals as the delta-functions, which is  justified as long as the characteristic wave-length of   doped carriers  significantly exceeds the orbital size $a_0$. A singular on-site (${\bf m}={\bf n}$) Coulomb repulsion of two carriers with the opposite spins (the Hubbard $U$) is infinite in this approximation. In fact, it should be cut  at $
\approx \alpha/\epsilon_{\infty}a_0$ as indicated by the bar above the sum, $\bar{\sum}$. Also for  mathematical transparency  we consider  a single electron band  dropping the electron band index.

Quantitative calculations of the EPI matrix elements in semiconductors and metals
 have to be performed numerically  from  pseodopotentials.
 Fortunately  one can parametrize EPI rather than  compute it  in
many physically important cases \cite{mahan}. EPI in  ionic lattices such as the cuprates is dominated by
coupling with polar optical phonons. This dipole interaction  is much stronger than the
deformation potential coupling to acoustic  phonons and  other multipole EPIs.  While  the EPI matrix
elements are ill-defined in metals,  they are well defined in doped
insulators, which have their parent dielectric compounds
with well-defined  phonon frequencies  $\omega _{{\bf q} }$ and the electron
band dispersion.

To parameterize EPI one can calculate the lowest order two-particle vertex function comprising the direct Coulomb repulsion and a phonon exchange \cite{mahan},
\begin{equation}
\Gamma({\bf q}, \Omega_n)={4 \pi \alpha\over{\epsilon_\infty V_0 q^2}}+|\gamma ({\bf q})|^{2}(\hbar \omega_{\bf q})^2 {\cal D}({\bf q}, \Omega_n)\;. \label{vertex}
\end{equation}
Here  ${\bf q}={\bf k}_1^\prime- {\bf k}_1$, $\Omega_n=\omega_{n^\prime1}-\omega_{n1}$ are the momentum and energy transfer in a scattering process of two carriers with the initial momenta and the Matsubara frequencies ${\bf k}_{1,2}$ and $\omega_{n1,2}$, respectively, and ${\cal D}({\bf q}, \Omega_n)=-\hbar\omega_{\bf q}/[\Omega_n^{2}+(\hbar\omega_{\bf
q})^{2}]$ is the propagator of a phonon of frequency $\omega_{\bf
q}$, and  $V_0$ is the unit cell volume. In the static limit, $\Omega_n=0$, Eq.(\ref{vertex}) yields the Fourier component of the particle-particle interaction as
\begin{equation}
\Gamma({\bf q}, 0)={4 \pi \alpha\over{\epsilon_\infty V_0 q^2}}-|\gamma ({\bf q})|^{2}\hbar \omega_{\bf q}. \label{int}
\end{equation}
 On the other hand, two static carriers localised on sites ${\bf m}$ and ${\bf n}$ in the ionic lattice repel each other with the Coulomb potential
\begin{equation}
v_{ij}={\alpha\over{\epsilon_{0}|\mathbf{%
m-n}|}}\;, \label{int2}
\end{equation}
where the static dielectric constant, $\epsilon_0$ accounts for the screening by both core electrons and ions. Comparing Eq.(\ref{int}) and Eq.(\ref{int2}) we find
\begin{equation}
|\gamma ({\bf
q})|^{2}\hbar\omega _{{\bf q}}={\alpha\over{\kappa}} \bar{\sum_{\bf m}}{e^{i{\bf q}\cdot{\bf m}}\over{m}} \approx {4\pi \alpha\over{\kappa V_0 q^2}},
 \label{gamma}
\end{equation}
 at relatively small $q \leq 1/a$. Here  $a$ is the lattice constant and
 $\kappa=\epsilon_0\epsilon_{\infty}/(\epsilon_0-\epsilon_{\infty})$ with the high-frequency dielectric constant $\epsilon_{\infty}$.
The static dielectric constant $\epsilon_{0}$ and the high-frequency dielectric constant $\epsilon _{\infty }$ are readily measured by putting the parent insulator in a capacitor and as the square of the refractive index of the insulator, respectively.
Hence, different from many models of high-temperature superconductors proposed so far, our generic Hamiltonian with the bare Coulomb and Fr\"ohlich interactions is defined through the measurable material parameters.

\section{$t-J_p$ and $t-J_p-\tilde{U}$ models}
  Using the Lang-Firsov (LF) canonical transformation \cite{LF} one can  integrate out most of \emph{%
both} interactions in the transformed Hamiltonian \cite{Alexandrov2011},
\begin{equation}
\tilde{H}=-\sum_{i,j}(\hat{\sigma}_{ij}\delta _{\sigma \sigma^{\prime }}+\tilde{\mu}%
\delta _{ij})c_{i}^{\dagger }c_{j}+H_{ph}+{\frac{1}{{2}}}\sum_{i\neq
j}v_{ij}n_{i}n_{j},  \label{trans}
\end{equation}%
since the residual repulsion, $v_{ij}$, is substantially  diminished by the large dielectric constant of the polar lattice [see Eq.(\ref{int2})].
Here
\begin{equation}
\hat{\sigma}_{ij}=T_{ij}\hat{X}_{i}^{\dagger }\hat{X}_{j} \label{sigmahat}
\end{equation}
is the renormalised hopping integral involving  multi-phonon transitions
 with $\hat{X}_{i}=\exp \left[ \sum_{\mathbf{q}}u_i(\mathbf{q})d_{%
\mathbf{q}}-H.c.\right] $ and $\tilde{\mu}$ is the chemical
potential shifted by the polaron binding energy.

Then using the  Schrieffer-Wolf (SW) canonical transformation \cite{sw} and neglecting $v_{ij}$ the transformed Hamiltonian Eq.(\ref{trans}) is reduced to the $t-J_p$ Hamiltonian as  \cite{Alexandrov2011}
\begin{eqnarray}
\mathcal{H}_{tJ_p}
&=& -\sum_{i,j}t_{ij}\delta_{\sigma\sigma^{\prime}}c_{i}^{\dagger}c_{j}+ \cr
&&2\sum_{\textbf{m}\neq\textbf{n}}J_{p}(\textbf{m}-\textbf{n})\left(\textbf{S}_\textbf{m}\cdot\textbf{S}_\textbf{n}+\frac{1}{4}n_{\textbf{m}}n_{\textbf{n}}\right)\;.
\label{eq:tJp_Hamiltonian}
\end{eqnarray}
Here the sum over $\textbf{n}\neq\textbf{m}$ counts each pair once only, $\textbf{S}_\textbf{m}=(1/2)\sum_{\sigma,\sigma^\prime}c^\dagger_{\textbf{m}\sigma}\overrightarrow{\tau}_{\sigma\sigma^\prime}c_{\textbf{m}\sigma^\prime}$ is the spin $1/2$ operator ($\overrightarrow{\tau}$ are the Pauli matrices), $n_{\textbf{m}}=n_{\textbf{m}\uparrow}+n_{\textbf{m}\downarrow}$, and $n_{\textbf{m}\uparrow, \downarrow}=c^\dagger_{\textbf{m}\uparrow, \downarrow}c_{\textbf{m}\uparrow, \downarrow}$ are site occupation operators.

All quantities in the
polaronic $t$-$J_{p}$ Hamiltonian \eqref{eq:tJp_Hamiltonian} are
defined through the  material parameters, in
particular the polaron hopping integral, $t_{ij}=T(%
\mathbf{m}-\mathbf{n})\exp[-g^2(\mathbf{m}-\mathbf{n})]$ with the polaron band-narrowing exponent
\begin{equation}
g^2(\mathbf{m})={\frac{2 \pi e^2}{{\kappa \hbar \omega_0
NV_0}}}\sum_\mathbf{q}{1-\cos(\mathbf{q }\cdot \mathbf{m})\over{q^2}},
\label{t}
\end{equation}
and
\begin{equation}
J_p(\mathbf{m})=T^2(\mathbf{m})/2g^2(\mathbf{m})\hbar\omega_0,\label{J}
\end{equation}

\begin{figure}[tbp]
\begin{center}
\includegraphics[width=0.6\columnwidth]{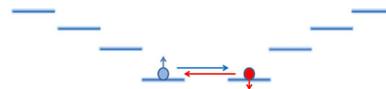}
\end{center}
\caption{(Color online) Exchange transfer of two polarons with opposite spins between nearest-neighbor sites with no potential barrier involved. Horizontal lines illustrate atomic levels shifted by the carrier-induced lattice deformation.}
\label{exchange}
\end{figure}

It has been proposed that the
$t$-$J_{p}$ Hamiltonian, Eq.(\ref{eq:tJp_Hamiltonian}), has a
high-T$_c$ superconducting ground state protected from clustering
\cite{Alexandrov2011}. The polaronic exchange $J_p$  is attractive for polarons in the singlet channel and repulsive for polarons in the triplet channel. The origin of this exchange attraction is illustrated in Fig.\ref{exchange}. If two polarons with opposite spins occupy nearest-neighbor sites, they can  exchange sites without any potential barrier between them, which lowers their energy by $J_p$ proportional to the unrenormalised hopping integral squared.

Importantly the LF trasfomation Eq.(\ref{trans}) is exact, and  the SW transformation is accurate for the intermediate and strong EPI coupling, $\lambda \geq 1/\sqrt {2z}$, where $\lambda$ is the BCS coupling constant and $z$ is the lattice coordination number as discussed in details in Ref. \onlinecite{Alexandrov2011}.
The residual repulsion of polarons, $v_{ij}$ in the transformed Hamiltonian, Eq.(\ref{trans}),  is small compared with the exchange inter-site polaron attraction $J_p$ and the short-range bipolaron-bipolaron repulsion of about the same magnitude, as long as $\epsilon_0 \gg \alpha/a J_p$. With the typical parameters of the cuprates  $J_p$ is about 1 eV and $\alpha/a \approx 4$ eV, so that the residual inter-site repulsion $v_{\textbf{mn}}$ is small if $\epsilon_0 \gg 1$, which is well satisfied in all relevant compounds \cite{alebra}.

\begin{figure}
\includegraphics[width=0.6\columnwidth]{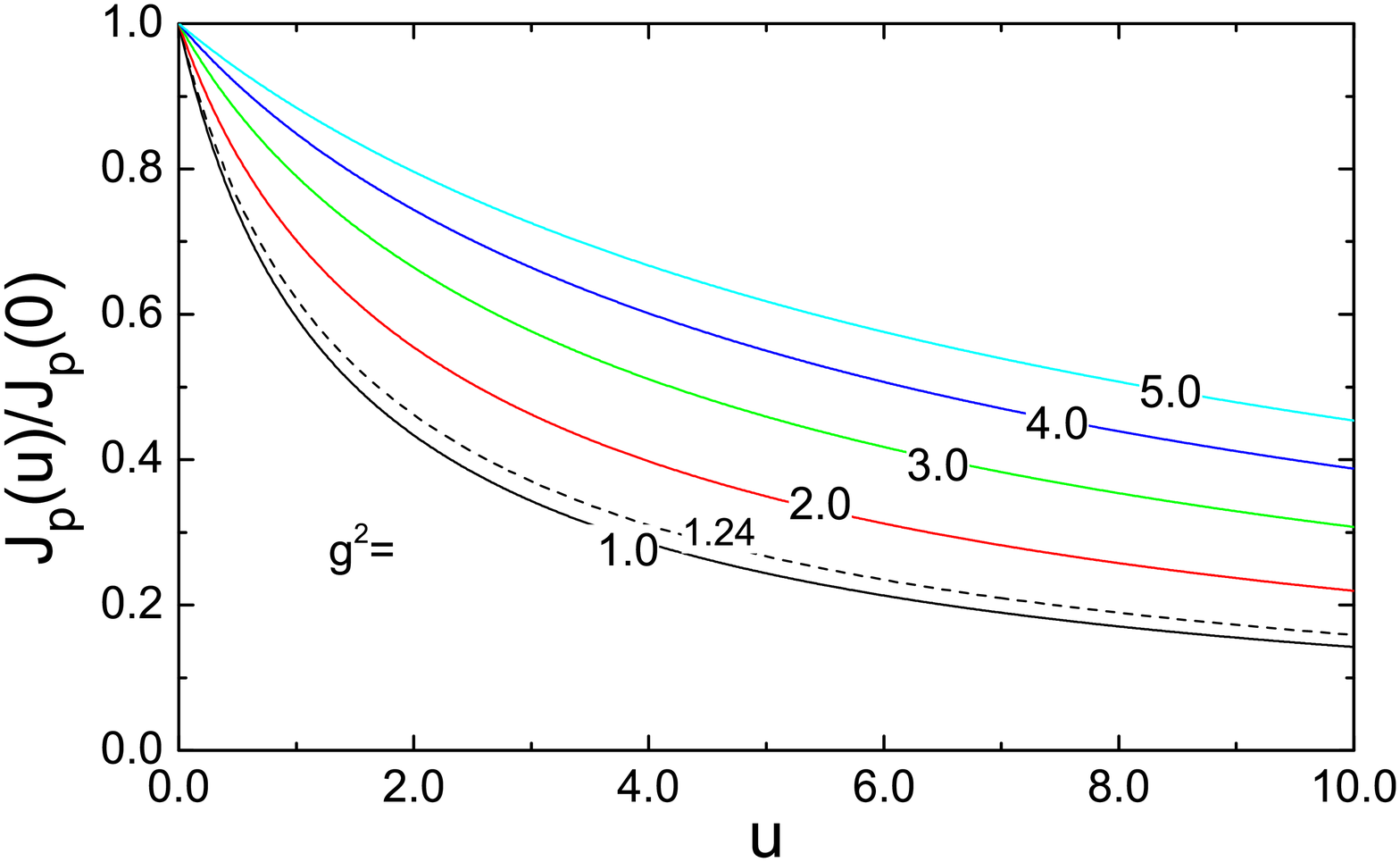}
\caption{(Color online) Reduction of the inter-site exchange attraction $J_p(u)/J_p(0)$ by the on-site residual polaron-polaron repulsion $u=\tilde{U}/\hbar \omega_0$ for different values of the polaron mass exponent $g^2$.} \label{u}
\end{figure}

Nevertheless the \textit{on-site} term in $v_{\textbf{mn}}$, Eq.(\ref{int2}), $\tilde{U}$ could be substantial, if the size of the Wannier orbitals is small enough $a_0 \ll a$. This renormalised $\tilde{U}$ is strongly di\-mi\-ni\-shed by the lattice polarization with respect to the bare on-site repulsion. We have emphasised in Refs.\onlinecite{Alexandrov2011,sica} that our model describes carriers doped into the charge-transfer Mott-Hubbard (or any polar) insulator, rather than the insulator itself, different from the conventional Hubbard U or t-J models.  The bare Hubbard-$U$ on the oxygen orbitals (where doped holes reside) in a rigid cuprate lattice is of the same order of magnitude as the on-site attraction induced by the Fr\"ohlich EPI ($\approx 1$ to $2$ eV \cite{alebra}), so that the residual Hubbard $\tilde{U}$ could be as large as a few hundred meV. We now take it into account  in the energy of a virtual double occupied state $\left|p\right\rangle$ with two opposite spins on the same site,
\begin{equation}
E_{p}-E_n =\tilde{U}+ \sum_{n_{\mathbf{%
q}}\neq 0}\hbar \omega_{\bf q} n_{\mathbf{%
q}}.
\end{equation}
Then performing the SW transformation the exchange attraction is found as
\begin{equation}
J_p(u,\textbf{m}-\textbf{n})=\frac{t^2}{\hbar\omega_0}\sum_{k=1}^\infty {(2g^2(\textbf{m}-\textbf{n}))^k\over{k! (k+u)}}\;,
\label{JU}
\end{equation}
where  $u=\tilde{U}/\hbar \omega_0$.  The reduction with respect to $J_p(0,\textbf{m})$ is moderate as long as the relative $u$ is less than $2g^2$, but becomes substantial for $u > 2g^2$, Fig.\ref{u}, which puts the characteristic bipolaron binding energy in the range of a hundred meV comparable with the double pseudogap in the cuprates \cite{aledev}. Importantly $J_p(u)$ remains large or comparable with the polaron hopping integral $t=T(\mathbf{a})\exp[-g^2(\mathbf{a})]$ since the spin exchange of the  $t-J_{p}$ model , Eq.(\ref{J}), does not contain the small polaron nar\-ro\-wing exponent $\exp (-g^2)$.

Hence our extended $t$-$J_{p}(u)$-$\tilde{U}$ model including major correlations effects  reads as follows
\begin{eqnarray}
\mathcal{H}
&=& -\sum_{i,j}t_{ij}\delta_{\sigma\sigma^{\prime}}c_{i}^{\dagger}c_{j} +\tilde{U} \sum_{\textbf{m}}n_{\textbf{m}\uparrow}n_{\textbf{m}\downarrow}+ \cr
&&+2\sum_{\textbf{m}\neq\textbf{n}}J_{p}(u,\textbf{m}-\textbf{n})\left(\textbf{S}_\textbf{m}\cdot\textbf{S}_\textbf{n}+\frac{1}{4}n_{\textbf{m}} n_{\textbf{n}}\right)\;.
\label{tJpU}
\end{eqnarray}

\section{Low density limit and high T$_c$}
As in Refs.\onlinecite{Alexandrov2011,sica} we adopt here the strong-coupling approach to the multi-polaron problem described by the Hamiltonian, Eq.(\ref{tJpU}), solving first a two-particle problem and then projecting the Hamiltonian on the repulsive Bose gas of small inter-site bipolarons. Such projection allows for a reliable estimate of the superconducting critical temperature for low carrier density as long as bipolarons remain small.

If we neglect the polaronic hopping taking $t=0$, then the ground and the highest energy states are bipolaronic spin-singlet and spin-triplet, respectively, made up of two
polarons on neighboring sites. The zero-energy states [in the nearest-neighbor (NN) approximation] are pairs of polarons separated by more than one lattice spacing. The on-site bipolaron has energy $\tilde{U}>0$.

\begin{figure}[tbp]
\includegraphics[width=1.0\columnwidth]{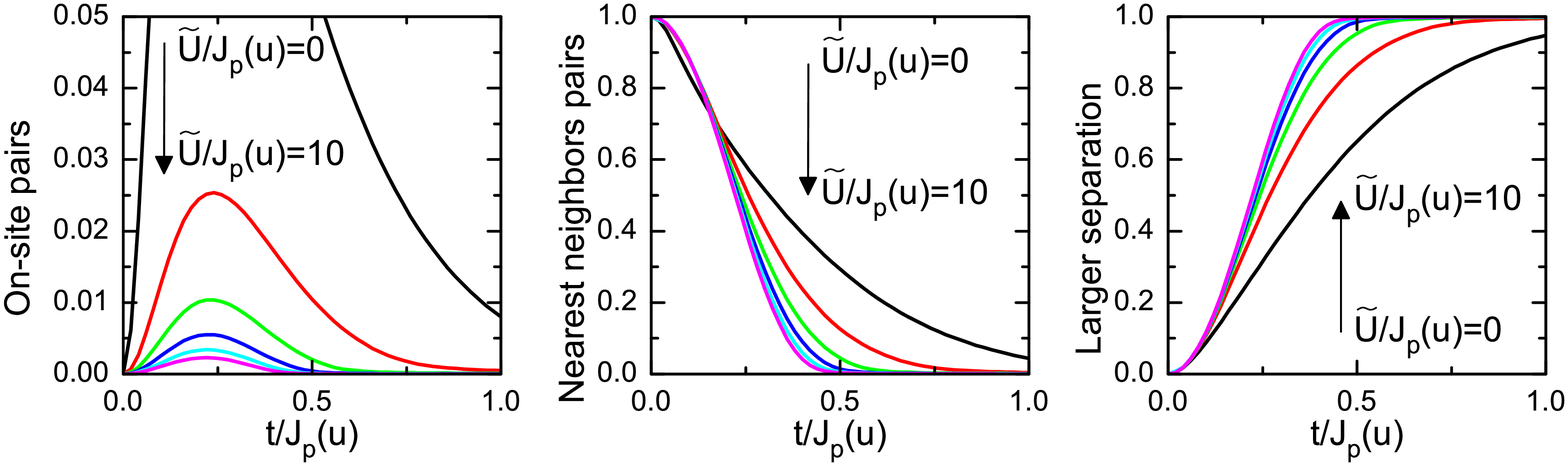}
\caption{(Color online) Probability to find two polarons on the
same site (left panel), on nearest-neighbor sites $P_{bp}$ (central panel), on more distant sites (right panel) in the ground state of the  $t-J_p(u)-\tilde{U}$ model on a $100\times100$ square lattice with different on-site repulsions.}
\label{prob}
\end{figure}

\begin{figure}[tbp]
\begin{tabular}{cc}
\includegraphics[width=0.5\columnwidth]{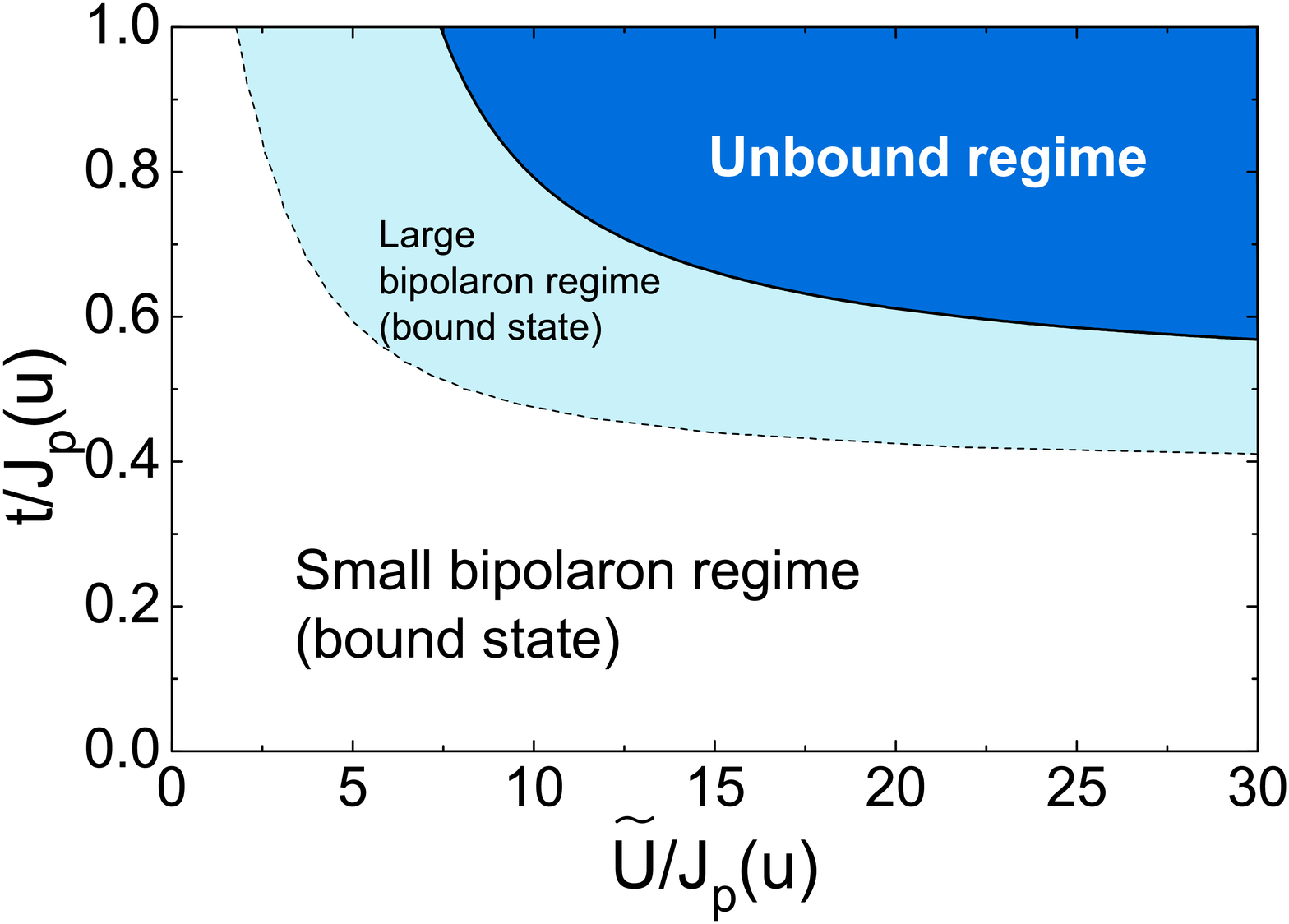}
\includegraphics[width=0.5\columnwidth]{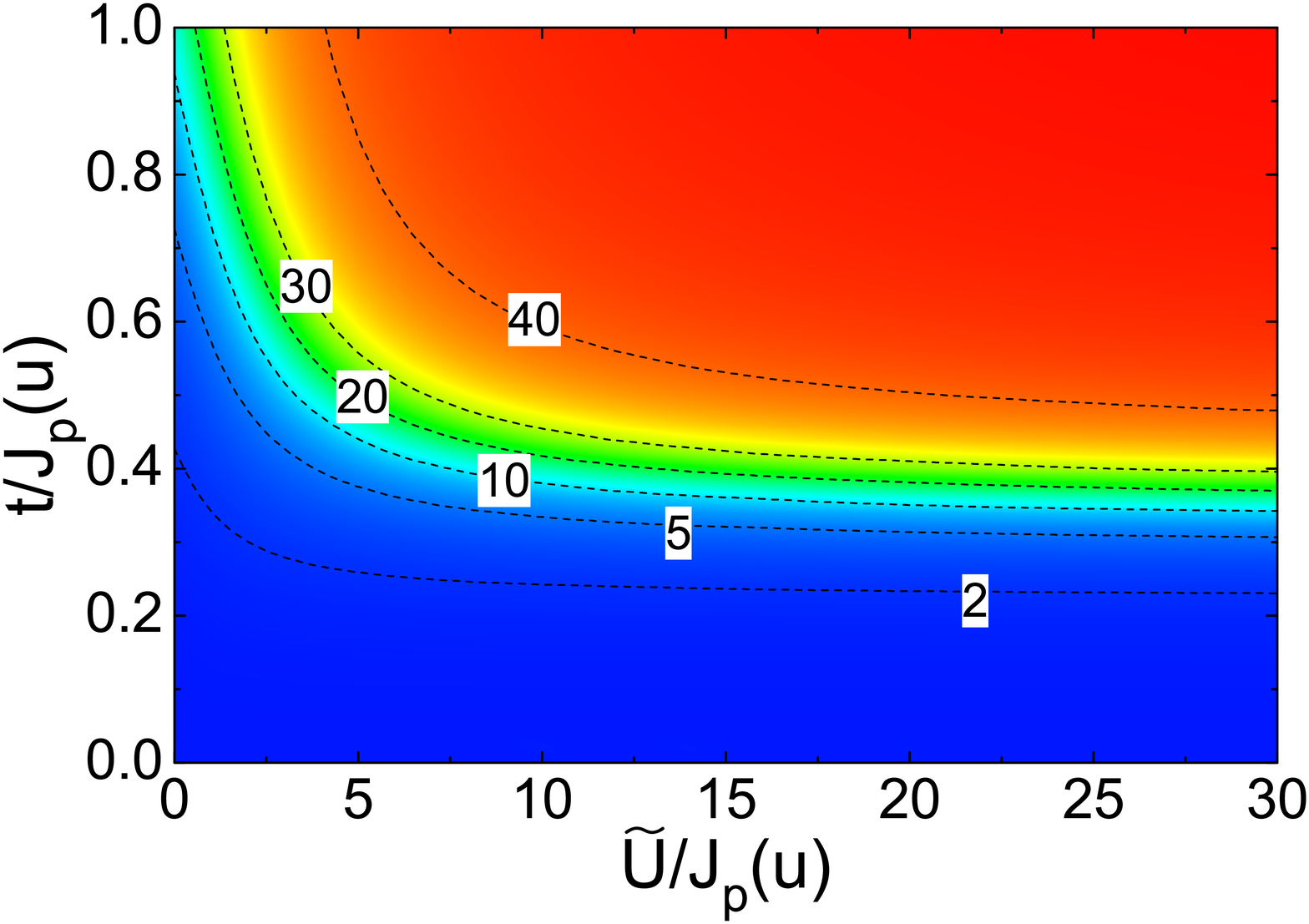}
\end{tabular}
\caption{(Color online) Left panel: phase diagram for the ground state of the polaronic $t$-$J_p(u)$-$\tilde{U}$ model on a square lattice. Right panel: contourplot of the bipolaron radius $r/a$ ($a$ is the lattice constant) for a $100\times100$ square lattice with periodic boundary conditions. Different numbers represent the value of $r/a$ along the boundaries (dashed lines), emphasizing the increasing of the bipolaron radius as we approach the unbound regime. Here $r=\langle |\textbf{m}-\textbf{n}|\rangle$, $\textbf{m}$ and $\textbf{n}$ being the position vectors of the two polarons in the bound state.}
\label{radius}
\end{figure}

\begin{figure}[tbp]
\includegraphics[width=0.65\columnwidth]{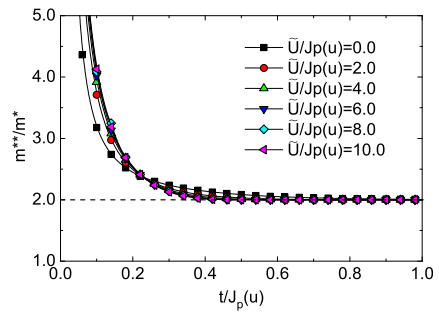}
\caption{(Color online) Ratio of bipolaron to polaron mass in the $t-J_p(u)-\tilde{U}$ model on a square lattice.}
\label{mass}
\end{figure}

For $t\neq0$ our exact diagonalization (ED) results on finite clusters show that the probability to find NN bipolarons falls as we increase the hopping or the strength of the on-site repulsion $\tilde{U}$ as shown in Fig.\ref{prob} for a $100\times100$ square lattice. Consistently, the bipolaron size increases but remains on the order of the lattice spacing in a wide domain of the parameters (see Fig.\ref{radius}). Importantly, although the small bipolaronic configuration persists for any values of the hopping at $\tilde{U}=0$, for $\tilde{U}\neq0$ and large values of $t$ up to a critical value $t_{c}=\tilde{U}J_p(u)/(2\tilde{U}-8J_p(u))$, the presence of a finite on-site interaction leads to the \emph{crossover from a small to a large bipolaronic configuration}. Finally, for further increasing $t$ the system undergoes a phase transition to an unbound state at $t=t_c$. The crossover from a small to a large bipolaronic configuration is also confirmed by the calculation of the bipolaron to polaron effective mass ratio with $m^{\ast\ast}=2m^\ast$ in the large bipolaron regime, as shown in Fig.\ref{mass}.

In the small bipolaron regime, the kinetic energy operator in Eq.(\ref{tJpU}) connects singlet configurations in the first and higher orders with respect to the polaronic hopping integrals. Taking into account only the singlet bipolaron band and discarding all other configurations one can map the $t-J_p(u)-\tilde{U}$ Hamiltonian on the hard-core charged Bose gas as described in Ref. \onlinecite{Alexandrov2011}. This gas is superfluid in 2D and higher dimensions. In particular, its 2D critical temperature $T_c$ in the dilute limit is given by \cite{Fisher1988}
\begin{equation}\label{Tcsuperfluid}
T_{c}={2\pi\hbar^{2}n_{b}\over{k_{B}m^{\ast
\ast}\ln\ln(1/n_{b}a^{2})}}\;,
\end{equation}
where $n_{b}$ is the boson density per unit area.

\begin{figure}[tbp]
\includegraphics[width=0.65\columnwidth]{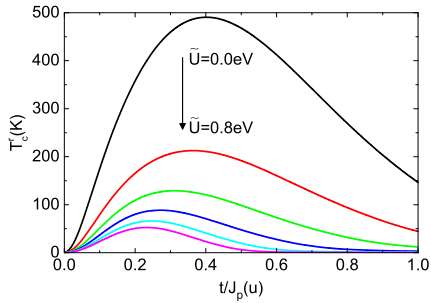}
\caption{(Color online) The superconducting critical tem\-pe\-ra\-tu\-re of the $t-J_p(u)-\tilde{U}$ model on the square lattice for low carrier density $n_b=0.05/a^2$ with $\hbar\omega_0=0.08$eV, $g^2=1.24$ and $J_p(0)=1$eV \cite{Alexandrov2011}.}
\label{Tc}
\end{figure}

The occurrence of superconductivity in this regime is not controlled by a pairing strength,
but by the phase coherence among small bipolarons \cite{AlexandrovRanninger}. At low enough density the Bose-Einstein condensation (BEC) temperature in 3D or its Berezinsky-Kosterlitz-Thouless (BKT) analog in 2D, Eq.(\ref{Tcsuperfluid}) should not significantly depend on the bipolaron size as long as it remains small. On the other hand increasing $\tilde{U}$ in our model finally results in a bipolaron overlap, where the bipolaron condensation should appear in the form of the \textit{polaronic} Cooper pairs in momentum space \cite{Alexandrov1983} with a lower critical temperature, rather than in real space (BEC-BCS crossover \cite{eagles,legget,Alexandrov1983,nozieres,mic,levin}). Hence, we can safely estimate the BEC critical temperature by weighting Eq.\ref{Tcsuperfluid} with the probability to find NN polarons as $T^{r}_{c}\approx P_{bp}(t/J_p(u))T_c$ \cite{sica}. As shown in Fig.\ref{Tc}, despite a low carrier density, for a physical choice of the parameters ($\hbar\omega_0=0.08$eV, $g^2=1.24$ and $J_p(0)=1$eV \cite{Alexandrov2011}) the cri\-ti\-cal temperature is found to be well in excess of $100$K for $\tilde{U}=0$ and rapidly decreases with increasing $\tilde{U}$.

It is worth noting that, unlike in other theories, the strength of the on-site interaction term reduces $T_c^r$. However, we recall that our residual on site interaction $\tilde{U}$ is defined as the difference between bare Hubbard $U$ and on-site Fr\"{o}hlich EPI therefore at $\tilde{U}=0$, when $T_c$ is maximized, we have a strong bare on-site interaction with $U\approx2E_p\sim 1$$-2$ eV.

\section{Conclusions}

In conclusion, we have introduced and studied the polaronic $t$-$J_p(u)$-$\tilde{U}$ model, defined through the bare material parameters. The model,  being an essential generalization of the $t$-$J_p$ model \cite{Alexandrov2011}, includes all electron-electron and electron-phonon correlations providing a microscopic explanation of the high-T$_c$ phenomenon without any ad-hoc approximations. We show that the inclusion of the residual on-site interaction $\tilde{U}$ (neglected in the $t$-$J_p$ model \cite{Alexandrov2011,sica}), drives the system to a BEC/BCS crossover that reconciles  the polaron-bipolaron theory of superconductivity with the observation of a large Fermi surface in overdoped cuprate superconductors. We offer an explanation, on microscopic grounds, of the high-T$_c$ phenomenon as a consequence of competing Coulomb and Fr\"ohlich interactions in highly polarizable ionic lattices beyond the conventional BCS description.

\section*{Acknowledgements}

We gratefully acknowledge enlightening discussions with Antonio Bianconi, Slaven Barisic, Ivan Bozovic, Victor Kabanov, Ferdinando Mancini, Dragan Mihailovic, Nikolay Plakida, and  support from the UNICAMP visiting professorship program and ROBOCON (Campinas, Brasil).
%\end{acknowledgments}

%%%%%%%%%%%%%%%%%%%%%%%%%%%%%%%%%%%%%%%%%%%%%%%%%%%%%%%%%%%%%%%%%%%%%%%%%%%%%
%   B I B L I O G R A P H Y
%%%%%%%%%%%%%%%%%%%%%%%%%%%%%%%%%%%%%%%%%%%%%%%%%%%%%%%%%%%%%%%%%%%%%%%%%%%%%

\end{document}